\documentclass[preprintnumbers,nofootinbib,showkeys,showpacs,amsmath,amssymb]{revtex4}
\usepackage{amsmath,amssymb,graphics,epsfig,subfigure}
\usepackage{color}
\usepackage{multirow}
\usepackage{booktabs}
\usepackage{changes}
\usepackage{footnote}

\usepackage{hyperref}
\hypersetup{colorlinks=true,linkcolor=blue,citecolor=magenta}

\begin{document}
	
\renewcommand{\baselinestretch}{1.3}

\title{Thermodynamical topology with multiple defect curves for dyonic AdS black holes}

\author{Zi-Qing Chen, Shao-Wen Wei \footnote{Corresponding author. E-mail: weishw@lzu.edu.cn}}

\affiliation{$^{1}$Lanzhou Center for Theoretical Physics, Key Laboratory of Theoretical Physics of Gansu Province, School of Physical Science and Technology, Lanzhou University, Lanzhou 730000, China\\
	$^{2}$Institute of Theoretical Physics and Research Center of Gravitation, 	Lanzhou University, Lanzhou 730000, People's Republic of China}

\begin{abstract}
Dyonic black holes with quasitopological electromagnetism exhibit an intriguing phase diagram with two separated first-order coexistence curves. In this paper, we aim to uncover its influence on the black hole thermodynamical topology. At first, we investigate the phase transition and phase diagram of the dyonic black holes. Comparing with previous study that there is no black hole phase transition region for a middle pressure, we find this region can narrow or disappear by fine tuning the coupling parameter. Instead, two first-order phase transitions can be observed. Importantly, we uncover that such novel phase diagram shall lead to a multiple defect curve phenomenon in black hole topology where each dyonic black hole is treated as one defect in the thermodynamical parameter space. By examining the topology, it is shown that there could be one, three, or five black hole states for given pressure and temperature. For each case, the topological number is calculated. Our results show that the topological number always takes value of +1, keeping unchanged even when the multiple defect curves appear. Therefore, our study provides an important ingredient on understanding the black hole thermodynamical topology.
\end{abstract}


\maketitle

\section{Introduction}

Black hole thermodynamics has been one of the active areas, and attracted much more attention among the black hole physics. Inspired by the anti-de Sitter/conformal field theory (AdS/CFT) correspondence \cite{acEWitten,acSRyu}, most recent study focuses on the extended phase space, where the cosmological constant is treated as the pressure \cite{Kastor}. Besides the small-large black hole phase transition \cite{Kubiznaka}, reminiscent of the liquid-gas phase transiton of the van der Waals (VdW) fluid, numerous intriguing phase transitions and structures have been discovered, such as reentrant phase transitions, triple points, and superfluid phases \cite{Altamiranoa,AltamiranoKubiznak,Dolan,Frassino,Cai,XuZhao,Tjoa2,Teob,Ruihong,Tavakoli}. These studies inspire the understanding of black hole phase transitions and the gravity-gauge duality \cite{Visser,Cong,Cong2,Ahmed,Gong}. Additionally, the microstructure of black holes, which offers insight into the dominant interactions among microscopic constituents, has been explored \cite{msGRuppiner1,msSWWei,msSWWei1,msSWWei2}.

Among the study of black hole thermodynamics, the topology has attracted great interest. There are two topological approaches aiming at uncovering the thermodynamical nature of different black hole systems. By introducing a vector with its zero points corresponding to the critical point, the topological concept was first proposed in Ref. \cite{toSWWEI}. In general, since the direction of the vector is ill-defined at its zero points, they can be treated as defects of the vector following Duan's pioneering $\phi$-mapping topological current theory \cite{YSD,YSD1}. From a topological perspective, each defect can be endowed with one winding number, taking integer values. Summing all of them shall give a topological number, which can be used to divide the systems into different topological classes sharing similar thermodynamical property in each class. In particular, according to the value of the winding number, the critical point is divided into two different classes, the conventional one and the novel one. They have opposite properties measuring the generation and annihilation of the black hole branches. Meanwhile, the black hole systems are cast in different class with different structures of the critical points. This study provides an initial exploration towards the black hole thermodynamical topology, and soon it was generalized to other black hole backgrounds \cite{tocYerra1,tocAhned,tocAlipour,tocGogoi1,tocChenZhang,tocBai,tocMYZ,tocZQC, tocAM1,tocFB}.

Although we can classify different black hole systems with the topology of the critical points, it is hard to realize for the reason that most well-known black holes do not possess the critical point. For example, both Schwarzschild black hole and charged Reissner-Nordstrom (RN) black hole have vanishing topological number, but their thermodynamical properties are quite different. The charged black hole can be thermodynamically stable or unstable, while the Schwarzschild black hole is always unstable. These stimulate us to consider the second construction of the black hole thermodynamical topology.

Let us go back to the beginning of the establishment of the black hole thermodynamics, the free energy is calculated by evaluating the Euclidean action in the form of
the gravitational path integral \cite{Gibbons}.  However, even for the simplest Schwarzschild black hole, it was found that negative heat capacity and imaginary energy fluctuations were produced. In order to cure these two shortcomings, York imagined that the black hole is placed inside a cavity with fixed temperature \cite{York}. Such treatment shall lead to that the free energy will be off-shell unless the temperature of the cavity equals the black hole Hawking temperature.

Employing with this generalized free energy, we proposed the second topology for the black hole thermodynamics \cite{topWei}. The vector was constructed by the off-shell free energy, and its zero points exactly corresponding to the actual black hole solutions. Detailed study shows that the local thermodynamical instability of the black holes is one kind topological property. These black holes with positive winding number are locally stable, otherwise they are unstable. Algebraically summing all the winding numbers shall give the total topological number. By making use of it, we conjecture that different black hole systems can be divided into three classes with number values -1, 0, and +1. The result showed that the Schwarzschild, charged RN, and charged RN-AdS black holes, respectively, have topological numbers -1, 0, and +1, indicating they are in different topological classes. In particular, this conjecture was partially confirmed in Ref. \cite{topWei}. Since the second treatment offers a well behaved topological approach, it was generalized to other black hole systems subsequently \cite{topDW1,topDW2,topDW3,topLiu,topFang,topFan,topDu,topYBD,topRL,topDYC1,topTNH,topCWT,topMR,topMYZ,topSPW}. The relation between the different phase transition types and the topological number was widely considered, including the ensemble and spacetime dimension.

On the other hand, the	quasi-topological electromagnetism term including the topological 4-form of Maxwell field gains interest \cite{qtAC,qtHSLiu,qtMHDehghani,qtRCMyer,qtYQLei,qtYSMyung,qtYZLi,qtAC1}, recently. Although the quasi-topological term does not contribute to the Maxwell equation and the energy-momentum tensor, it contributes significantly to the dyonic black hole solutions, and many novel features emerge. There could be four horizons and three photon spheres \cite{qtHSLiu}. The issues, chaotic behaviors and particle motion were considered in Ref. \cite{qtYQLei}. It indicates that the chaos bound will be violated if the quasitopological electromagnetism term is included in. In particular, the thermodynamics and phase transition were studied in Ref. \cite{qtLMDa}, where a triple point was observed. Moreover, a novel phase structure with two separated coexistence curves was discovered, which is quite different from the charged RN-AdS black hole case with only the small-large black hole phase transition.

It has been revealed that for the VdW-like phase transition and triple point, the defect curve, a set of the zero point of the vector, is continuous. This probably results the conservation of the topological number for the black holes. For the dyonic black holes, since the separated coexistence curves appear, the defect curve may not be one single curve. This presents a novel topological pattern for the black hole thermodynamics. Inspired by it, we aim to uncover the underlying topological property of the dyonic black hole thermodynamics, and study whether the topological number remains unchanged even when the multiple defect curves are presented.

The present paper is organized as follows. In Sec. \ref{sec:thermodynamics}, we briefly review the dyonic black hole thermodynamics. The phase transition and phase structure are studied in Sec. \ref{sec:COEXISTENCE}, where two separated coexistence curves are observed. In particular, three fine phase structures are exposed. In Sec. \ref{sec:TOPOLOGY}, following the approach of Ref. \cite{topWei}, we construct the topology for the dyonic black holes. Under the multiple defect curve case, we calculate the topological number. Our result shows that the topological number still remains unchanged, providing an important ingredient to the black hole thermodynamical topology. Finally, the paper is concluded with a summary and discussion of our results in Sec. \ref{Sec_Disscusion}.
	
\section{Thermodynamics of dyonic black holes}\label{sec:thermodynamics}

In this section, we would like to give a brief review of the thermodynamics of the dyonic black holes.

In four dimensions, the Lagrangian of dyonic black holes can be expressed as \cite{qtHSLiu}
\begin{eqnarray}
 \mathcal{L}=\sqrt{-g}(R-2\Lambda)+\alpha_1\mathcal{L}_{\text{M}}+\alpha_2\mathcal{L}_{\text{Q}},\label{eq:Lagrangian}
\end{eqnarray}
which includes the minimally coupling of the Maxwell electromagnetism and quasitopological electromagnetism. Here, $\Lambda$ represents the cosmological constant. The coupling constant $\alpha_1$ is dimensionless and $\alpha_2$ has dimension of $[\text{length}]^{2}$. The standard Maxwell Lagrangian $\mathcal{L}_{\text{M}}$ and the quasitopological electromagnetic Lagrangian $\mathcal{L}_{\text{Q}}$ are given by
	\begin{eqnarray}
		\mathcal{L}_{\text{M}}&=&-\sqrt{-g}F^2,\label{eq:Lagrangian2}\\
		\mathcal{L}_{\text{Q}}&=&-\sqrt{-g}\left((F^2)^2-2F^{(4)}\right),\label{LQ} \label{eq:Lagrangian1}
	\end{eqnarray}
	where $F^2=-F^\mu{}_\nu F^\nu{}_\mu$, $F^{(4)}=F^\mu{}_\nu F^\nu{}_\rho F^\rho{}_\sigma F^\sigma{}_\mu$, and the Maxwell field tensor reads $F_{\mu\nu}=\partial_{\mu}A_{\nu}-\partial_{\nu}A_{\mu}$ with $A_{\mu}$ the vector potential.
	
The black hole solution in the static and spherically symmetric spacetime background can be obtained by solving the corresponding Einstein field equations, given by \cite{qtHSLiu}
\begin{eqnarray}
	ds^2&=&-f(r) dt^2+\frac{1}{f(r)} dr^2+r^2 (d\theta^2+\sin^2\theta d\phi^2),\\
	f(r)&=&1-\frac{2M}{r}+\frac{\alpha_1 p^2}{r^2}-\frac{1}{3}\Lambda r^2+\frac{q^2}{\alpha_1r^2}~_{2}F_1[\frac{1}{4},1;\frac{5}{4};-\frac{4p^2\alpha_2}{r^4\alpha_1}].
\end{eqnarray}
Here, the parameters $p$ and $q$ are related to the electric charge $Q_{\rm e}$ and magnetic charge $Q_{\rm m}$ of the black holes
\begin{eqnarray}
	Q_{\rm e}=\frac{1}{4\pi}\int\tilde{F}^{0r}=q, \quad
	Q_{\rm m}=\frac{1}{4\pi\alpha_1}\int F=\frac{p}{\alpha_1}. \label{Qm}
\end{eqnarray}
Following Ref. \cite{Kastor}, the cosmological constant $\Lambda$ can be treated as the pressure
\begin{eqnarray}
	P=-\frac{1}{8 \pi} \Lambda.
\end{eqnarray}
Employing with the ``Euclidean trick", the Hawking temperature can be calculated as
\begin{eqnarray}
T=\frac{f^{\prime}(r_{h})}{4 \pi}=\frac{1}{4\pi r_{\rm h}}+2P r_{\rm h}-\frac{Q_{\rm m}^2 \alpha_1^3}{4\pi r_{\rm h}^3}-\frac{Q_{\rm e}^2r_{\rm h}}{4\pi(r_{\rm h}^4\alpha_1+4Q_{\rm m}^2\alpha_1^2\alpha_2)}.\label{temp}
\end{eqnarray}
The black holes entropy can be obtained by the Beckenstein-Hawking area-entropy formula
\begin{eqnarray}
	S&=&\pi r_{\rm h}^2.\label{eq:entropy}
\end{eqnarray}
By solving $f(r_{h})=0$, the black hole mass can be expressed as
\begin{eqnarray}
	M&=&\frac{3r_{\rm h}^2\alpha_1+8P\pi r_{\rm h}^4\alpha_1+3Q_{\rm m}^2\alpha_1^4+3Q_{\rm e}^2~_{2}F_1[\frac{1}{4},1;\frac{5}{4};-\frac{4Q_{\rm m}^2\alpha_1\alpha_2}{r_{\rm h}^4}]}{6r_{\rm h}\alpha_1}.\label{eq:mass}
\end{eqnarray}

It is well-known that the global thermodynamic stability of a black hole is determined by its free energy, as a thermodynamic system naturally tends towards the state of lowest free energy. Therefore, the Gibbs free energy is a crucial quantity for studying the phase transitions, capable of exhibiting swallowtail behavior indicative of a first-order phase transition within the system. The expression for Gibbs free energy can be derived by evaluating the Euclidean action through the gravitational path integral approach. Within the ``zero-loop" approximation, the partition function is expressed as follows
\begin{eqnarray}
	\mathcal{Z}=e^{- \beta G }=\int_{ }^{ }D[g]e^{- \frac{\mathcal{S}}{\hbar} } \sim e^{- \frac{\mathcal{S}}{\hbar} },
\end{eqnarray}
where $\mathcal{S}$ is the Euclidean action, and Gibbs free energy is given by
\begin{eqnarray}
	G=-\frac{1}{\beta} \ln\mathcal{Z}=\frac{\mathcal{S}}{\beta}.\label{eq:g}
\end{eqnarray}
Here $\beta$ is the inverse of the black hole temperature. Utilizing Eq. \eqref{eq:Lagrangian}, the bulk action $\mathcal{S}_{bulk}$ reads
\begin{eqnarray}
	\mathcal{S}_{bulk}=\frac{1}{16 \pi} \int_{ }^{ } dx^{4} \left( \sqrt{-g}(R-2\Lambda)+\alpha_1\mathcal{L}_{\text{M}}+\alpha_2\mathcal{L}_{\text{Q}} \right).\label{eq:sb}
\end{eqnarray}
In order to obtain a well-behaved action, one needs to introduce the Gibbons-Hawking-York boundary term and counter term
\begin{eqnarray}
	\mathcal{S}_{GH}=\frac{1}{16 \pi} \int_{ \partial M}^{ } dx^{3}  \sqrt{-\gamma}2K ,\qquad \mathcal{S}_{ct}=\frac{1}{16 \pi} \int_{\partial M }^{ } dx^{3} \sqrt{-\gamma} \left(\frac{4}{l} + l R^{(\partial M)}\right),
\end{eqnarray}
where $\sqrt{-\gamma}$ is the induced metric and $K$ is the extrinsic curvature on the boundary $\partial M$. AdS radius $l$ corresponds to the cosmological constant, and $R^{(\partial M)}$ is curvature of boundary $\partial M$. Combining with them, the total action $\mathcal{S}$ shall be
\begin{eqnarray}
	\mathcal{S}=\mathcal{S}_{bulk}-\mathcal{S}_{GH}-\mathcal{S}_{ct}.
\end{eqnarray}
Substituting it into Eq.\eqref{eq:g}, we get
\begin{eqnarray}
	G&=&\frac{Q_e^2 \,
		_2F_1\left(\frac{1}{4},1;\frac{5}{4};-\frac{4 Q_m^2
			\alpha _1 \alpha _2}{r_h^4}\right)}{2 \alpha _1
		r_h}+\frac{Q_e^2 r_h^3}{4 \alpha _1 \left(r_h^4+4
		\alpha _1 \alpha _2 Q_m^2\right)}+\frac{3 \alpha _1^3
		Q_m^2}{4 r_h}-\frac{2}{3} \pi  P r_h^3+\frac{r_h}{4}.
\end{eqnarray}
Via examining this Gibbs free energy, the thermodynamical phase transition and phase structure can be obtained.

\section{Separated coexistence curve}\label{sec:COEXISTENCE}

In Ref. \cite{qtLMDa}, the thermodynamics phase transitions and coexistence curves have been examined. Besides the VdW-like phase transition and triple point phase diagrams, there presents another novel interesting one, the separated coexistence curve phase diagram. For small and large pressures, there exists one first-order phase transition. But in the middle pressure region, no phase transition can be observed. Here we wonder whether there exist more fine phase structures. For simplicity, we set the electric charge, magnetic charge, and coupling constant to $Q_{e}=5$, $Q_{m}=2.5$, and $\alpha _{2}=50$. Moreover, we take three distinct cases $\alpha _{1}$=0.3, 0.406831, and 0.41 for examples.

\subsection{Case I: $\alpha_{1}=0.3$}\label{sec:0.4G}

\begin{figure*}[h]
	\centering
	\subfigure[]{\includegraphics[scale=0.4]{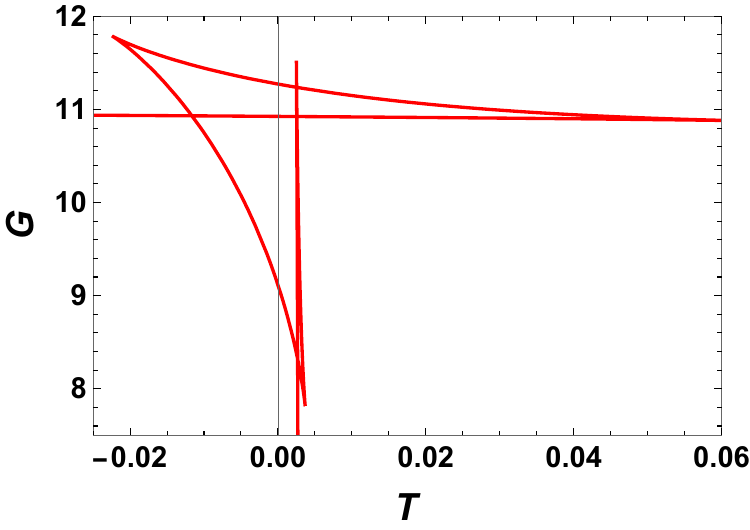}\label{fig:0.4GT1}}
	\subfigure[]{\includegraphics[scale=0.4]{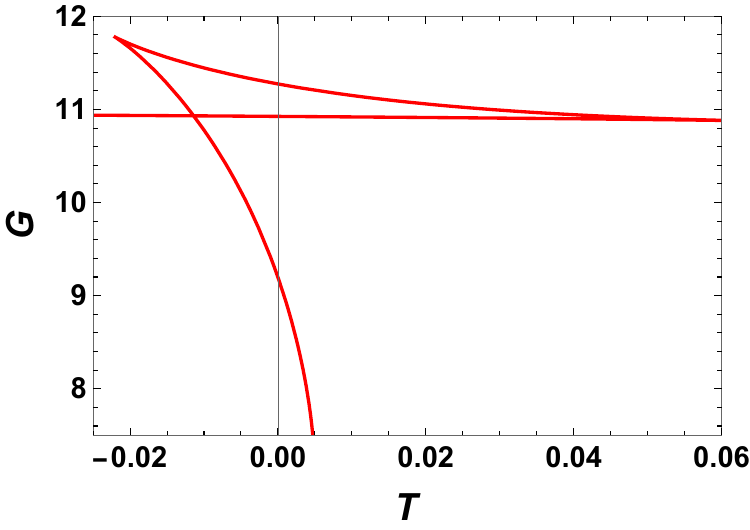}\label{fig:0.4GT2}}
	\subfigure[]{\includegraphics[scale=0.4]{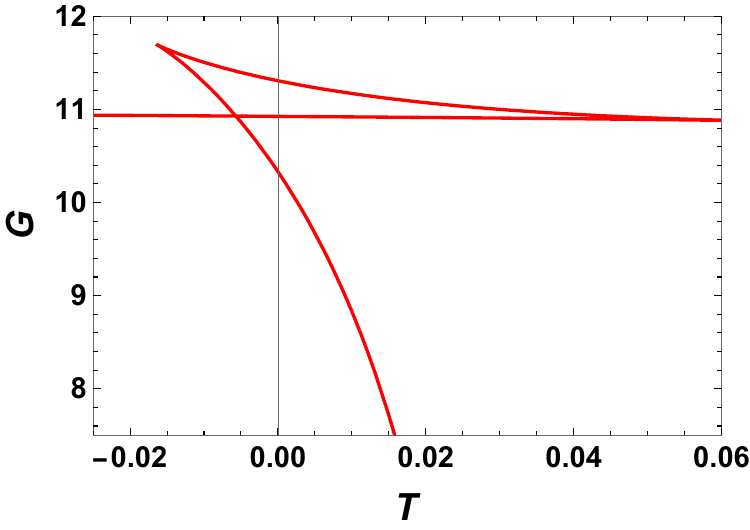}\label{fig:0.4GT3}}
	\subfigure[]{\includegraphics[scale=0.4]{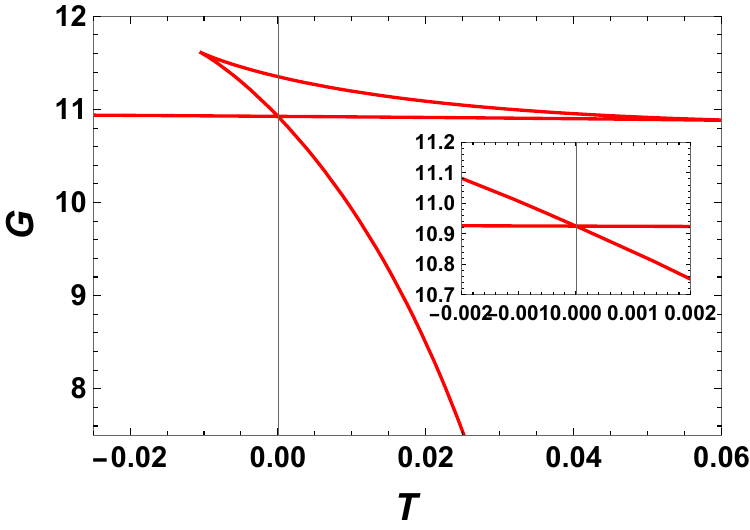}\label{fig:0.4GT4}}
	\subfigure[]{\includegraphics[scale=0.4]{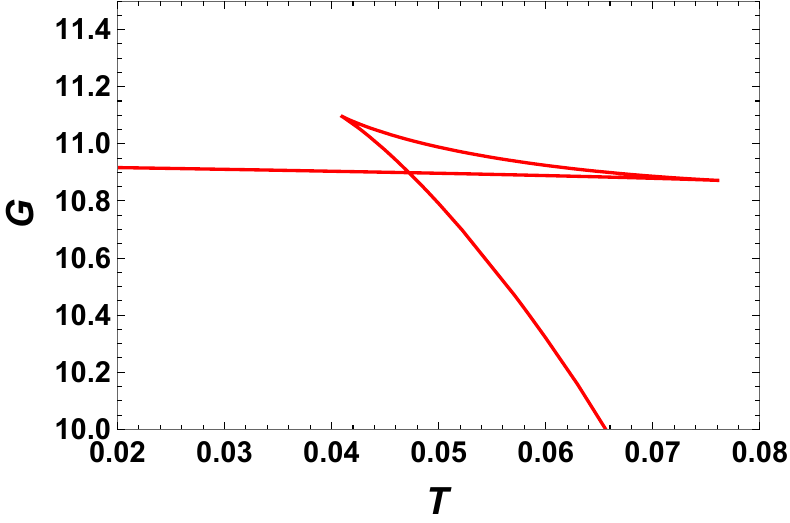}\label{fig:0.4GT5}}
	\subfigure[]{\includegraphics[scale=0.4]{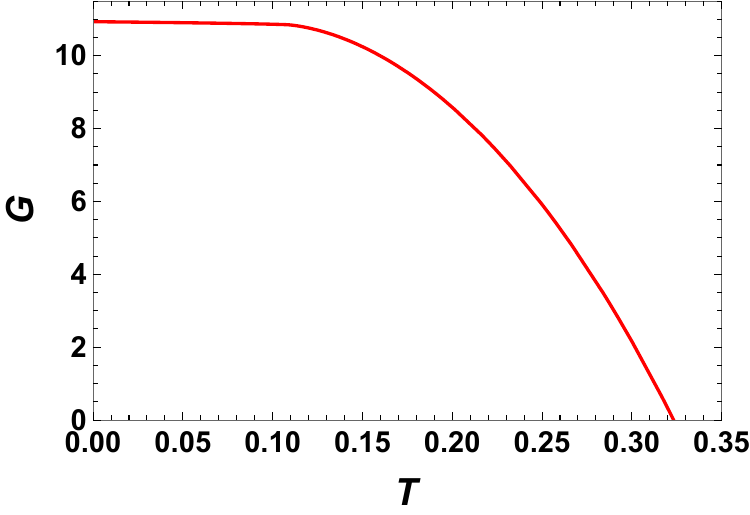}\label{fig:0.4GT6}}
	\caption{Behaviors of the Gibbs free energy $G$ with respect to the temperature $T$ with $\alpha_{1}=0.3$ and $\alpha_{2}=50$. With the increase of pressure $P$, the number of swallowtail behaviors gradually decrease from two to zero. (a)$P=0.00001$. (b)$P=P_{c1}=0.000040$. (c)$P=0.0008$. (d)$P=P_{0}=0.001605$. (e)$P=0.01$. (f)$P=0.03$.}\label{fig:0.4GT}
\end{figure*}

Let us first focus on $\alpha_{1}=0.3$. By solving
\begin{eqnarray}
	\left(\frac{\partial T}{\partial r_{h}}\right)_{P}=0, \quad \left(\frac{\partial^{2} T}{\partial r_{h}^{2}}\right)_{P}=0,
\end{eqnarray}\label{eq:Trhpartial}
one obtains two critical points
\begin{eqnarray}
	T_{c1}=0.004747, \quad P_{c1}=0.000040.\\
	T_{c2}=0.105254, \quad P_{c2}=0.028371.
\end{eqnarray}\label{eq:0.4criticalPT}
Near them, we study the behaviors of the Gibbs energy $G$. The results are exhibited in Fig. \ref{fig:0.4GT}. When the pressure $P < P_{c1}$, two swallowtail behaviors are presented, as displayed in Fig. \ref{fig:0.4GT1}. However, one of them locates in the negative temperature region. Thus it is unphysical and we abandon it. As a result, only the another swallowtail behavior is left, indicating an intermediate-large black hole phase transition. When $P = P_{c1}$, the swallowtail behavior with a positive temperature disappears. For this case, there is no phase transition due to the fact that the remaining swallowtail behavior still has a negative temperature, see Fig. \ref{fig:0.4GT2} and Fig. \ref{fig:0.4GT3}. As one increases the pressure $P$ gradually to $P = P_{0} = 0.001605$, as shown in Fig. \ref{fig:0.4GT4}, the temperature of the intersection point of the remaining swallowtail behavior equals zero. Continuing to increase the pressure $P$, it leads to a new phase transition, indicating a small-intermediate transition. Finally, by setting $P > P_{c2}$, the Gibbs free energy $G$, described in Fig. \ref{fig:0.4GT6}, becomes monotonic function of the temperature, which suggests that the system undergoes no phase transition.

\begin{figure}[h]
	\centering
	\subfigure[]{\includegraphics[scale=0.55]{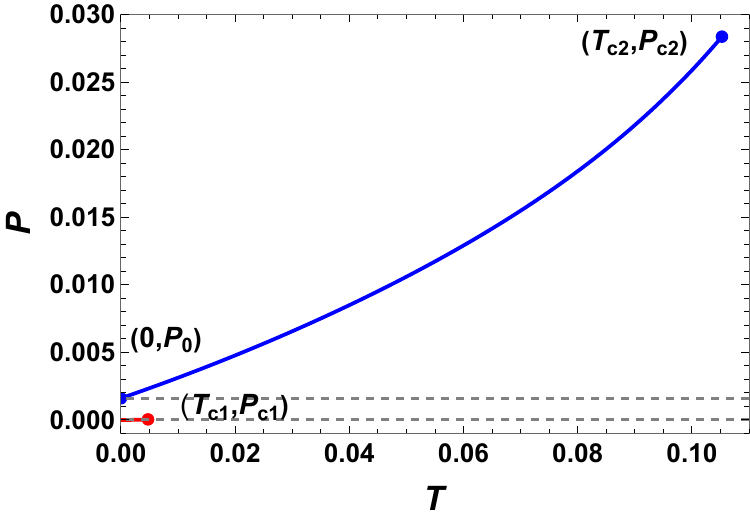}\label{fig:0.4PT1}}
	\subfigure[]{\includegraphics[scale=0.6]{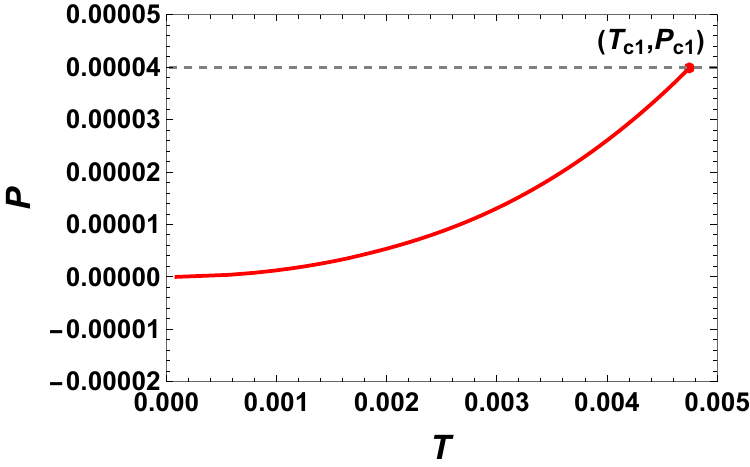}\label{fig:0.4PT2}}
	\caption{Phase diagram for the dyonic AdS black hole with $\alpha_{1}=0.3$ and $\alpha_{2}=50$. (a) Phase diagram. (b) The zooming phase diagram near $(T_{c1},P_{c1})$.}\label{fig:0.4PT}
\end{figure}

Through examining the behavior of the Gibbs free energy for different values of the pressure, we illustrate the $P$-$T$ phase diagram of the dyonic AdS black hole in Fig. \ref{fig:0.4PT}. As depicted, this phase diagram markedly differs from that of the VdW fluid. Two separated coexistence curves appear and they do not intersect with each other. One begins at the origin and ends at $(T_{c1}, P_{c1})$, while the other starts on the $P$ axis at $(0, P_{0})$ and ends at $(T_{c2}, P_{c2})$. In particular, for $P_{c1}<P<P_{0}$, the phase transition behaviors vanish. However, in the regions where $P < P_{c1}$ and $P_{0} < P < P_{c2}$, intermediate-large and the small-intermediate phase transition behaviors are observed, respectively. This peculiar feature, not seen in RN-AdS black holes, signifies a novel phase structure unique to dyonic AdS black holes with quasitopological electromagnetism, and this result is exactly consistent with that of Ref. \cite{qtLMDa}.

\subsection{Case II: $\alpha_{1}=0.406831$}
\label{sec:0.406831G}

From above subsection, we have observed the separated coexistence curves of the phase transition. It also exhibits that, for the middle pressure, there is no phase transition. Here, we are interested in whether there exists a value of the coupling constant $\alpha_{1}$, which can make such no phase transition region disappear. Fortunately, we find that, with the increase of $\alpha_{1}$, this region narrows. In particular, when $\alpha_{1}$ increases to 0.406831, the region disappears. Here we shall consider the details of this case.

The critical points for this case read
\begin{eqnarray}
	T_{c1}=0.005528, \quad P_{c1}=0.000054.\\
	T_{c2}=0.066054, \quad P_{c2}=0.012401.
\end{eqnarray}\label{eq:0.4069criticalPT}
Obviously, the temperature and pressure of the second critical point are still far beyond the first one.

\begin{figure*}[h]
	\centering
	\subfigure[]{\includegraphics[scale=0.4]{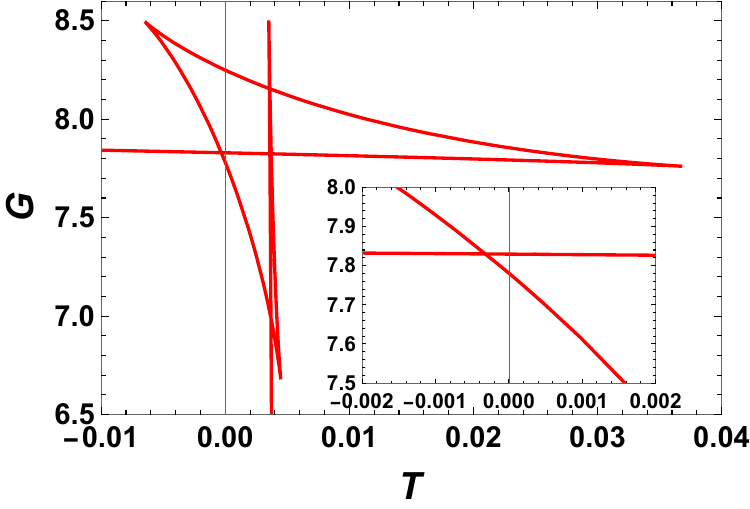}\label{fig:0.4069GT1}}
	\subfigure[]{\includegraphics[scale=0.4]{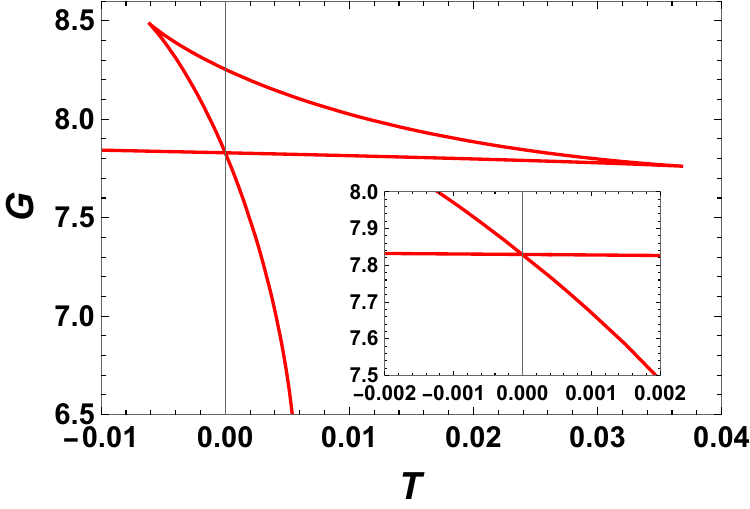}\label{fig:0.4069GT2}}
	\subfigure[]{\includegraphics[scale=0.4]{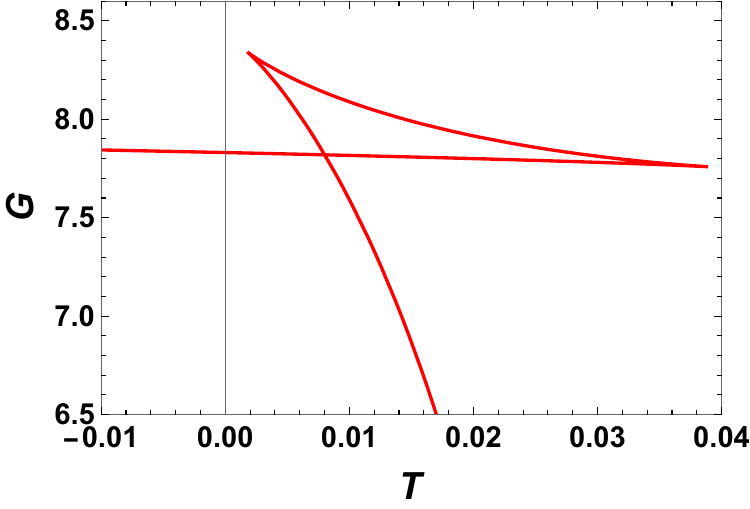}\label{fig:0.4069GT3}}
	\caption{Behaviors of the Gibbs free energy $G$ with respect to the temperature $T$ with $\alpha_{1}=0.406831$ and $\alpha_{2}=50$. (a) $P=0.00002$. (b) $P=P_{0}=P_{c1}=0.000054$. (c) $P=0.001$.}\label{fig:0.4069GT}
\end{figure*}

The detailed behaviors of the Gibbs free energy $G$ are plotted in  Fig. \ref{fig:0.4069GT} for  $P=0.00002$, $0.000054$ ($P_{0}$), and $0.001$, respectively. In Fig. \ref{fig:0.4069GT1}, the pressure is smaller than that of the first critical point. We can see that there are two swallowtail behaviors. One of them has negative temperature, and thus only the intermediate-large black hole phase transition can occur. For the second case, the pressure is tuned exactly to $P=P_{c1}=P_{0}$. From Fig. \ref{fig:0.4069GT2}, we observe an interesting phenomenon, the right side swallowtail behavior tends to disappear, meanwhile the intersection point of the left swallowtail behavior has zero temperature at the same $P$. This suggests that we have one first-order phase transition at zero temperature and one second-order phase transition at $T_{c1}$ for the same pressure.

\begin{figure}[h]
	\centering
	\subfigure[]{\includegraphics[scale=0.6]{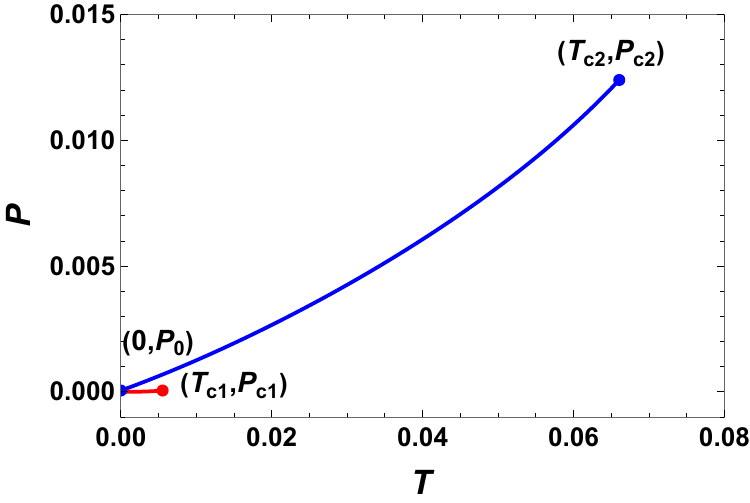}\label{fig:0.4069PT1}}
	\subfigure[]{\includegraphics[scale=0.6]{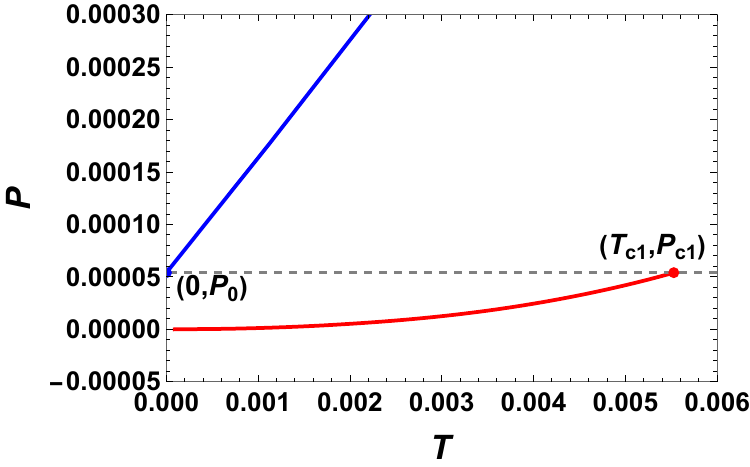}\label{fig:0.4069PT2}}
	\caption{Phase diagram for the dyonic AdS black hole with $\alpha_{1}=0.406831$ and $\alpha_{2}=50$. (a) Phase diagram. (b) The zooming phase diagram near $(T_{c1},P_{c1})$.}\label{fig:0.4069PT}
\end{figure}

In order to clearly show the case, we exhibit the phase diagram in Fig. \ref{fig:0.4069PT}. From the figure, it is easy to see that the no phase transition region just disappear.

\subsection{Case III: $\alpha_{1}=0.41$}
\label{sec:0.41G}

As shown above, the no phase transition region just disappears by increasing the coupling constant such that $\alpha_{1}=0.406831$. It is worth examining what shall happen for a larger $\alpha_{1}$. For such purpose, we set $\alpha_{1}=0.41$.

\begin{figure*}[h]
	\centering
	\subfigure[]{\includegraphics[scale=0.4]{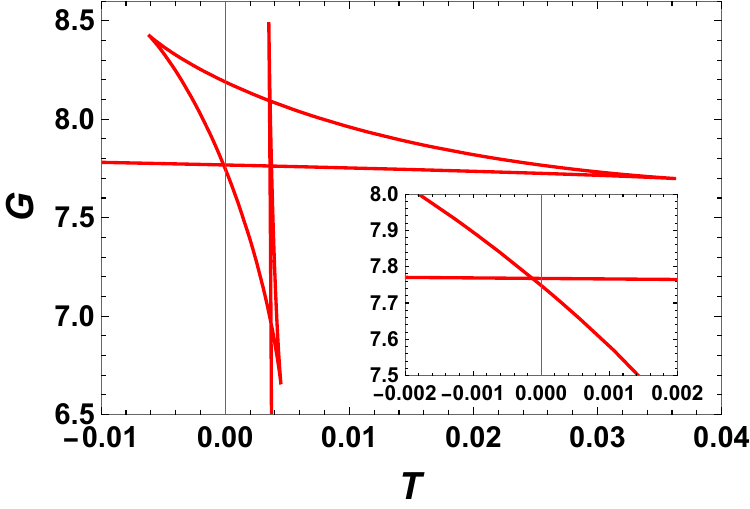}\label{fig:0.41GT1}}
	\subfigure[]{\includegraphics[scale=0.4]{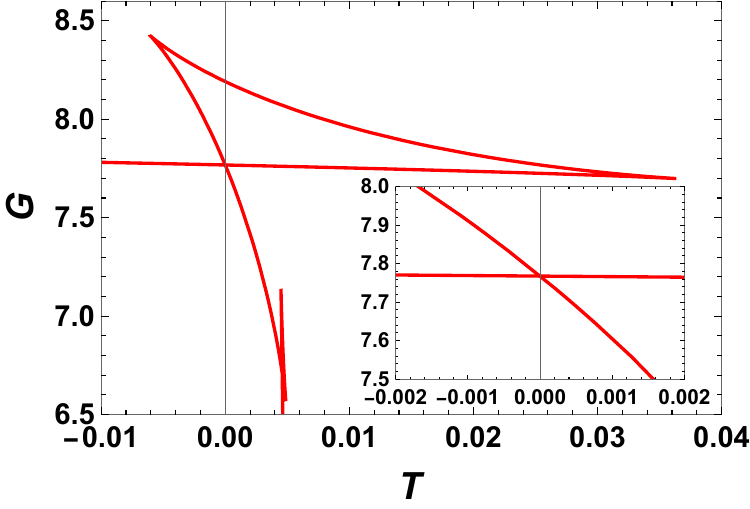}\label{fig:0.41GT2}}
	\subfigure[]{\includegraphics[scale=0.4]{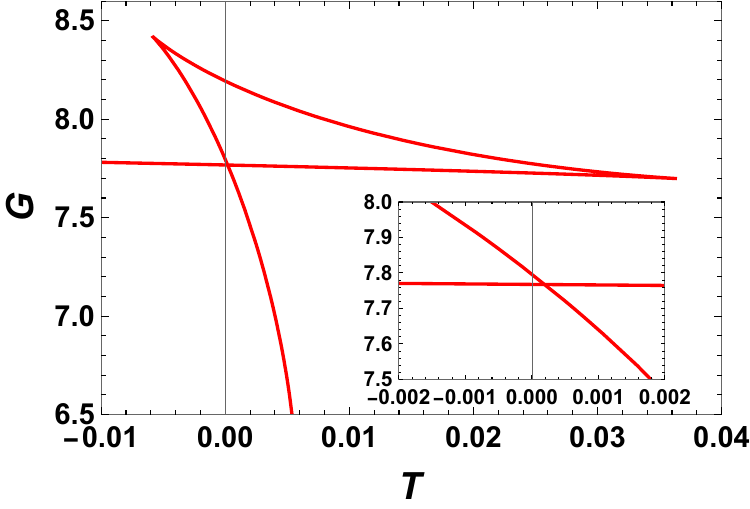}\label{fig:0.41GT3}}
	\caption{Behaviors of the Gibbs free energy $G$ with respect to the temperature $T$ with $\alpha_{1}=0.41$ and $\alpha_{2}=50$. (a) $P=0.00002$. (b) $P=P_{0}=0.000034$. (c) $P=P_{c1}=0.000054$.}\label{fig:0.41GT}
\end{figure*}

For this case, there are still two critical points
\begin{eqnarray}
	T_{c1}=0.005528, \quad P_{c1}=0.000054.\\
	T_{c2}=0.012401, \quad P_{c2}=0.012401.
\end{eqnarray}
The behavior of the Gibbs free energy is shown in Fig. \ref{fig:0.41GT} for different values of the pressure. For small pressure shown in Fig. \ref{fig:0.41GT1}, the left swallow tail behavior do not participate in the phase transition. So the right one gives the intermediate-large black hole phase transition. Slight increasing the pressure such that $P=P_{0}=0.000034$, the left swallow tail behavior gives a phase transition at zero temperature, see Fig. \ref{fig:0.41GT2}. Further increasing the pressure, it shall give a small-intermediate black hole phase transition. As a result, two phase transitions can take place for the middle pressure. As the pressure tends to its first critical value, the right swallow tail behavior disappears, leaving the small-intermediate black hole phase transition shown in Fig. \ref{fig:0.41GT3}.

\begin{figure}[h]
	\centering
	\subfigure[]{\includegraphics[scale=0.6]{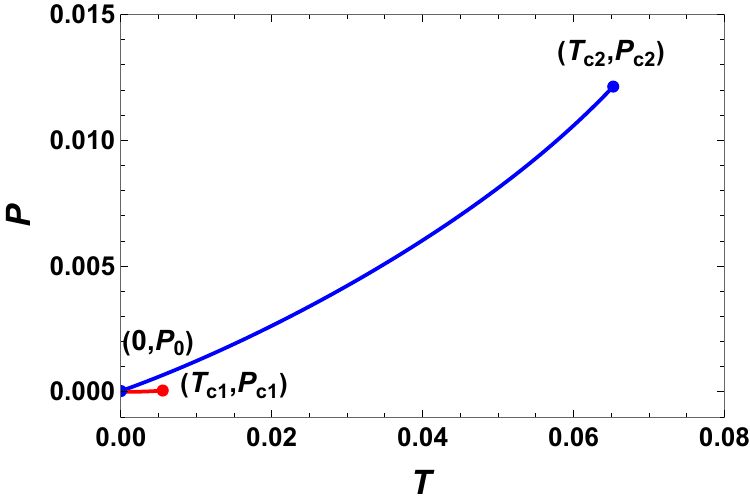}\label{fig:0.41PT1}}
	\subfigure[]{\includegraphics[scale=0.6]{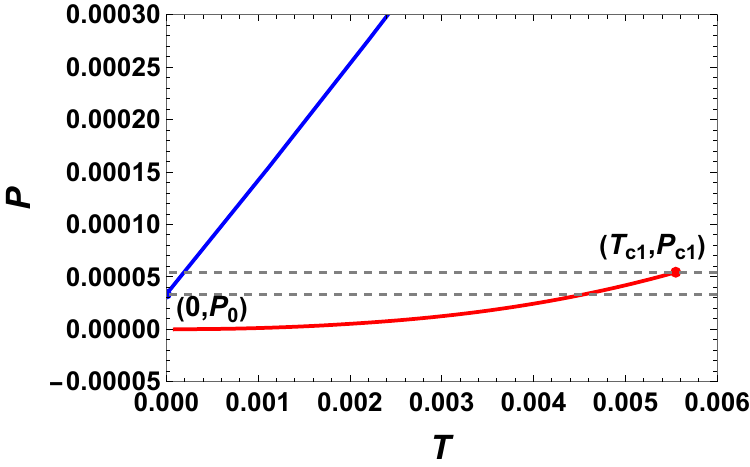}\label{fig:0.41PT2}}
	\caption{Phase diagram for the dyonic AdS black hole with $\alpha_{1}=0.41$ and $\alpha_{2}=50$. (a) Phase diagram. (b) The zooming phase diagram near $(T_{c1},P_{c1})$.}\label{fig:0.41PT}
\end{figure}

The phase diagram is exhibited in Fig. \ref{fig:0.41PT}. For small or large pressure, we have only one intermediate-large or small-large black hole phase transition. While in the middle pressure region, there are two phase transitions at two different temperatures. This strongly suggests that the no phase transition region completely disappears.

In this section, we uncover the detailed phase diagram including two separated coexistence curves for the dyonic black holes. Besides such novel phase diagram pattern, it is worth pointing out that there are some states having negative temperature, see the left swallow tail behavior of the Gibbs free energy. Abandoning these unphysical system states, we shall see in the next section that the defect curves shall be multiple. This naturally rises one question whether does the topological number get changed or remain unchanged.

\section{Topology and multiple defect curves}
\label{sec:TOPOLOGY}

In this section, following the topological approach proposed in Ref. \cite{topWei}, we would like to study the thermodynamical topology by treating the dyonic black holes as defects in the thermodynamical parameter space. In particular, we aim to investigate the topology nature when the black hole system exhibits the separated coexistence curves.

Considering the black hole thermodynamics, one can introduce the generalized off-shell free energy
\begin{eqnarray}
\mathcal{F}=M-\frac{S}{\tau},\label{eq:offshellfreeenergy}
\end{eqnarray}
for a black hole thermodynamical system with the mass $M$ and the entropy $S$. The parameter $\tau$ is an extra variable that can be regarded as the inverse temperature of the cavity surrounding the black hole. The on-shell free energy shall be recovered if $\tau=1/T$.

In order to establish a well behavior topology, we can introduce the following vector $\phi$ \cite{topWei}
\begin{eqnarray}
 \phi=\left(\frac{\mathcal{ \partial F}}{\partial r_{h}},-\cot \Theta \csc \Theta\right), \label{eq:vectorphi}
\end{eqnarray}
where the two parameters $r_{h}$ and $\Theta$ obey $0<r_{h}<+\infty$ and $0<\Theta<\pi$, respectively. Since the component $\phi^{\Theta}$ is divergent at $\Theta=0$, $\pi$, the direction of the vector points outward there. For such construction, the black hole solution is exactly located at the zero point of the vector and is limited at $\Theta=\pi/2$.

According to Duan's $\phi$-mapping topological current theory \cite{YSD,YSD1}, the corresponding topological current is defined as
\begin{eqnarray}
	j^{\mu}=\frac{1}{2 \pi} \epsilon^{\mu \nu \rho} \epsilon_{a b} \partial_{\nu} n^{a} \partial_{\rho} n^{b}, \quad \mu, \nu, \rho=0,1,2,
\end{eqnarray}\label{eq:topologicalcurrent}
where $n^{a}=\frac{\phi^{a}}{\|\phi\|}$ is the unit vector, and $x^{\mu}$=($\tau$, $r_{h}$, $\Theta$). It is easy to check that this topological current is conserved, i.e., $\partial_{\mu}j^{\mu}=0$. By making use of the Jacobi tensor, the two-dimensional Laplacian Green function, and $\delta$-function theory, the topological current takes the following simple form
\begin{equation}
	j^{\mu}=\delta^{2}(\phi)J^{\mu}\left(\frac{\phi}{x}\right).\label{juu}
\end{equation}
Significantly, it is easy to see that $j^{\mu}$ is nonzero only at the zero points of $\phi$. Denoting its $i$-th solution as $\vec{x}=\vec{z}_{i}$, the density of the topological current reads
\begin{equation}\label{24}
	j^{0}=\sum_{i=1}^{N} \beta_{i} \eta_{i} \delta^{2}\left(\vec{x}-\vec{z}_{i}\right),
\end{equation}
where the Hopf index $\beta_{i}$ counts the number of the loops that $\phi^{a}$ makes when $x^{i}$ goes around the $i$-th zero point counterclockwise, and the Brouwer degree $\eta_{i}$ denotes the sign of $J^{0}\left(\phi / x \right) _{z_{i}}$. For a given parameter region $\Sigma$, the corresponding topological number can be obtained via integrating the density of the topological current
\begin{equation}
	W=\int_{\Sigma} j^{0} d^{2} x=\sum_{i=1}^{N} \beta_{i} \eta_{i}=\sum_{i=1}^{N} w_{i},\label{25}
\end{equation}
with $w_{i}$ denoting the winding number of the $i$-th zero point of $\phi$. So it is clear that the topological number is the algebraic sum of the winding numbers of the zero points enclosing in the considered region $\Sigma$. Moreover, for two regions containing the same zero points, they shall have the same topological number. If the region does not contain any zero point, this number vanishes.

From Eqs. (\ref{eq:entropy}) and (\ref{eq:mass}), the generalized off-shell free energy reads
\begin{eqnarray}
	\mathcal{F}=\frac{\alpha_{1} p^{2}}{3 r_{h}}+\frac{r_{h}}{2}+ \frac{4 P \pi r_{h}^{3}}{3}-\frac{\pi  r_{h}^{2}}{\tau}+\frac{q^{2}  {}_{2}F_{1}\left(\frac{1}{4},1,\frac{5}{4},-\frac{4\alpha_{2} p^{2}}{\alpha_{1} r_{h}^{4}}\right)}{3\alpha_{1}r_{h}},
\end{eqnarray}\label{eq:offshellfreeenergydyonic}
and the components of the vector $\phi$ can be calculated as
\begin{eqnarray}
	\phi^{r_{h}}&=&\frac{1}{2}-\frac{\alpha_{1} p^{2}}{2 r_h^2}+4\pi P r_h^2-\frac{2 \pi r_h}{\tau}-\frac{q^2 r_h^2}{8 \alpha_{2} p^2 + 2 \alpha_{1} r_h^4},\label{pppp}\\
	\phi^{\Theta}&=&-\cot \Theta \csc \Theta.
\end{eqnarray}\label{eq:component}
Next, we plan to consider the topological number and defect curves for the dyonic black hole with $\alpha_{1}=0.41$ and $\alpha_2=50$. For simplicity, we take the pressure $P r_{0}^{2}=$0.05, 0.0003, and 0.00004 as three characteristic examples. Here $r_{0}$ is an arbitrary length scale set by the size of a cavity surrounding the black hole.

\subsection{Case I: $P r_{0}^{2}$=0.05}

For this case, the pressure is beyond its critical value, and thus no phase transition occurs.

\begin{figure}
	\centering{
	\subfigure[]{\includegraphics[width=6cm]{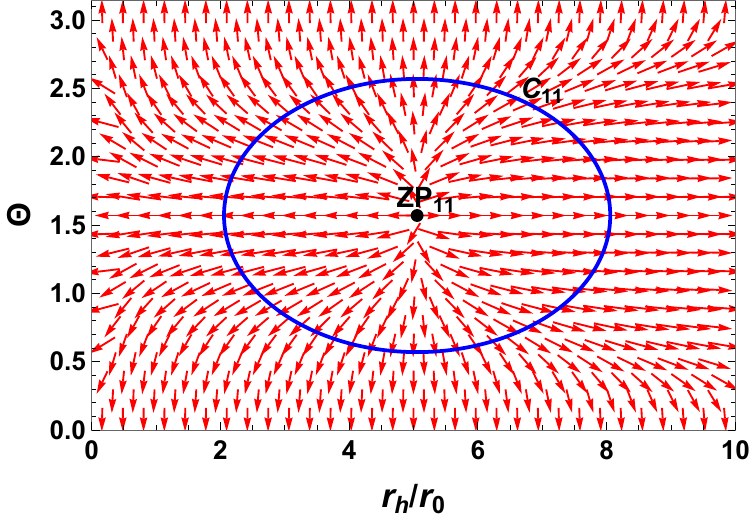}\label{pCavector}}
	\subfigure[]{\includegraphics[width=6cm]{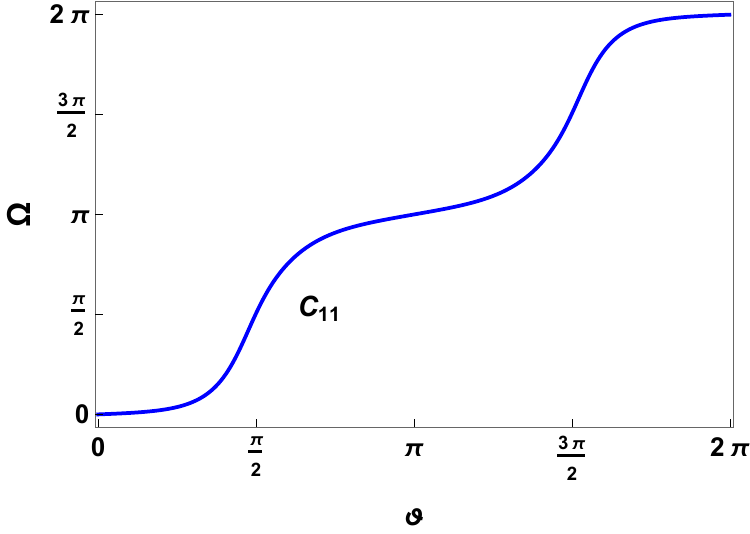}\label{pCatnumber}}}
	\caption{(a) The unit vector field on the $r_h/r_0$-$\Theta$ plane with $P r_0^2$ = 0.05 and $\tau/r_0$ = 2. The zero point $ZP_{11}$ located at ($r_{h}/r_{0}$, $\Theta$)=($5.0557$, $\pi/2$) which is marked with black dot, and it is enclosed with the blue loop $C_{11}$. (b) The deflection angle $\Omega$ as a function of $\vartheta$ for the contour $C_{11}$.}\label{ppCatnumber}
\end{figure}

We plot the unit vector in Fig. \ref{pCavector} with $\tau /r_0=2$. From it, we can see that there is only one zero point locating at ($r_{h}/r_{0}$, $\Theta$)=($5.0557$ , $\pi/2$). At this point, the direction of the vector is ill-defined, and thus it acts as a defect in this parameter space. Actually, this point is just related with the dyonic black hole. Near $\Theta=\pi/2$, the vector direction is to the left for $r_{h}/r_{0}<5.0557$, while to the right when $r_{h}/r_{0}>5.0557$. At $\Theta=0$ and $\pi/2$, it is obvious that the vector is outward.

In order to obtain the wind number of the zero point, we need to construct a closed loop around it, and then count the changes of the vector direction. For the purpose, we parametrize the closed loop C by the angle $\vartheta$ as
\begin{eqnarray}
\left\{
\begin{aligned}
 r_h/r_0&=a\cos\vartheta+r_c, \\
 \Theta&=b\sin\vartheta+\frac{\pi}{2}.
\end{aligned}
\right.\label{pfs}
\end{eqnarray}
Varying $\vartheta$ from $0$ to $2\pi$, one shall go counterclockwise around the zero point along this closed loop once. In order to count the change of the direction of the vector, it is convenient to define the deflection angle \cite{Wei:2020rbh}
\begin{equation}
 \Omega(\vartheta)=\oint_C \epsilon_{ab}n^a\partial_{i}n^bdx^{i}.
\end{equation}
Then the winding number shall be

\begin{equation}
	w=\Omega(2\pi)/2\pi.
\end{equation}
We choose $(a, b, r_c)$=$(3, 1, 5.0557)$ for $C_{11}$. Then the deflection angle can be calculated. The result is shown in Fig. \ref{pCatnumber}. It is easy to see $\Omega(\vartheta)$ increases monotonically with $\vartheta$. When $\vartheta$ tends to $2\pi$, $\Omega$ also takes value of $2\pi$. This implies that the winding number corresponding to the zero point $w=1$. Since there is only one zero point, the total topological number $W=1$. It suggests that the dyonic black hole for this case is in the same topological class of the charged RN-AdS black hole, while different from the Schwarzschild black hole. On the other hand, the sign of the winding number also indicates the local stability of the black hole. So we have one locally thermodynamically stable black hole at $\tau /r_0=2$.

\begin{figure}
\includegraphics[width=6.5cm]{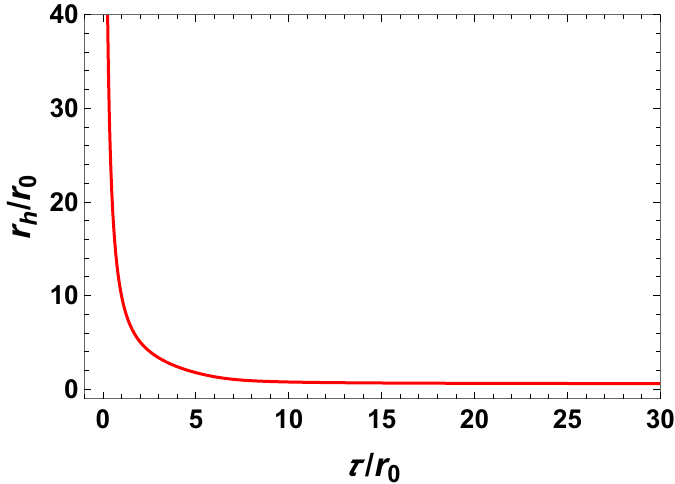}
\caption{Defect curve in the $r_h/r_0-\tau/r_0$ plane with $P r_0^2$ = 0.05.}\label{pCataur}
\end{figure}

Now, let us turn to the defect curve, a set of zeros of the vector, in the parameter space. This defect curve is easy to obtain by solving Eq. (\ref{pppp}). We depict the result in $r_{h}/r_{0}-\tau/r_{0}$ in Fig. \ref{pCataur}. With the increase of $\tau/r_{0}$, the radius $r_{h}/r_{0}$ of the black hole horizon corresponding to the defect of the vector decreases monotonically. A detailed calculation shows that all the defects or zero points along this curve have winding number +1, which strongly demonstrates this result is independent of $\tau$. Meanwhile, we observe only one single defect curve. This phenomenon is quite similar to the previous cases, such as the Schwarzschild black hole, charged black hole and so on, ignoring the topological number.

\subsection{Case II: $P r_{0}^{2}$=0.0003}

In the previous case, we see one local thermodynamical stable black hole branch corresponding the defects with winding number +1. Here we decreases the pressure such that $P r_{0}^{2}$=0.0003. In what followings, we shall see there will present another novel feature for the defects and defect curves.

\begin{figure}
	\centering{
	\subfigure[]{\includegraphics[width=6cm]{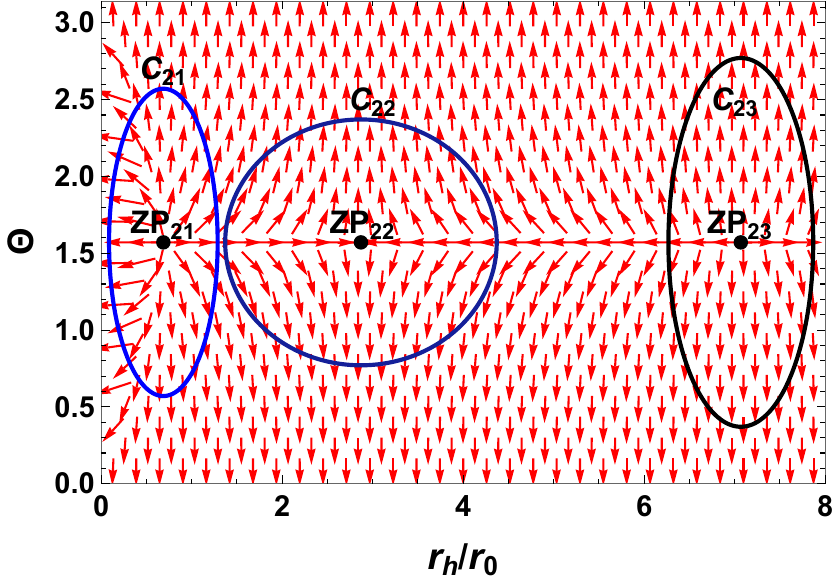}\label{Cbvector}}
	\subfigure[]{\includegraphics[width=6cm]{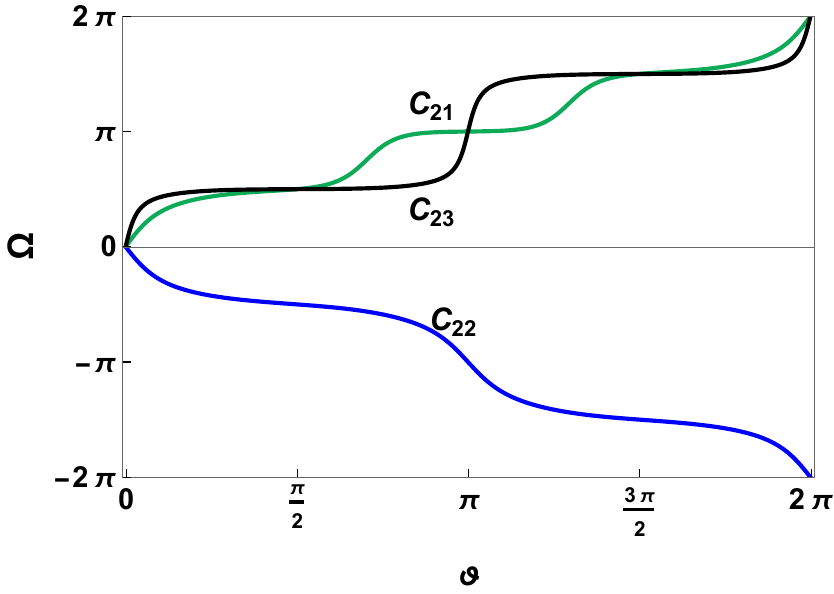}\label{Cbtonumb}}}
	\caption{(a) The unit vector field on the $r_h/r_0$-$\Theta$ plane with the $P r_0^2$ = 0.0003 and $\tau/r_0$ = 250. The zero points $ZP_{21}$, $ZP_{22}$, and $ZP_{23}$ located at ($r_{h}/r_{0}$, $\Theta$)=($0.6869$ , $\pi/2$), ($2.8703$ , $\pi/2$), and ($7.0674$ , $\pi/2$) which are marked with black dots, and they are enclosed by the loops $C_{21}$, $C_{22}$ and $C_{23}$, respectively. (b) The deflection angle $\Omega$ as a function of $\vartheta$ for contours $C_{21}$ (green), $C_{22}$ (blue), and $C_{23}$ (black).}\label{ppCatnumber}
\end{figure}

Taking $\tau /r_0=250$, we plot the unit vector $n$ on the ($r_{h}/r_{0}$, $\Theta$) plane in Fig. \ref{Cbvector}. Contrary to the first case, we observe three zero points located at $r_h/r_0$=0.6869, 2.8703, and 7.0674. Along $\Theta=\pi/2$, the vector turns its direction once it crosses a zero point. In particular, along the boundary $\Theta=0$, $\pi$, $r_{h}=0$, and $\infty$, both cases I and II share the similar behavior of the vector, indicating that they have the same topological number, and which will be confirmed soon.

By constructing three closed loops $C_{21}$, $C_{22}$, and $C_{23}$, respectively marked in green, blue, and black curves, to encircle these zero points, one is allowed to calculate the winding number for each defect. Note that these loops are parameterized with the form given in Eq. (\ref{pfs}), and with the parameters $(a, b, r_c)$=(0.6, 1, 0.6869), (1.5, 0.8, 2.8703), and (0.8, 1.2, 7.0674). In order to clear show it, we illustrate the deflection angle in Fig. \ref{Cbtonumb}. From the behavior, we can see that the zero points $ZP_{21}$ and $ZP_{23}$ have winding number +1, while $ZP_{22}$ has -1. Summing these winding numbers, we obtain the topological number $W=+1-1+1=+1$. Therefore, for $\tau /r_0=250$, we find the topological number of case II is the same with case I.

\begin{figure}
\includegraphics[width=6.5cm]{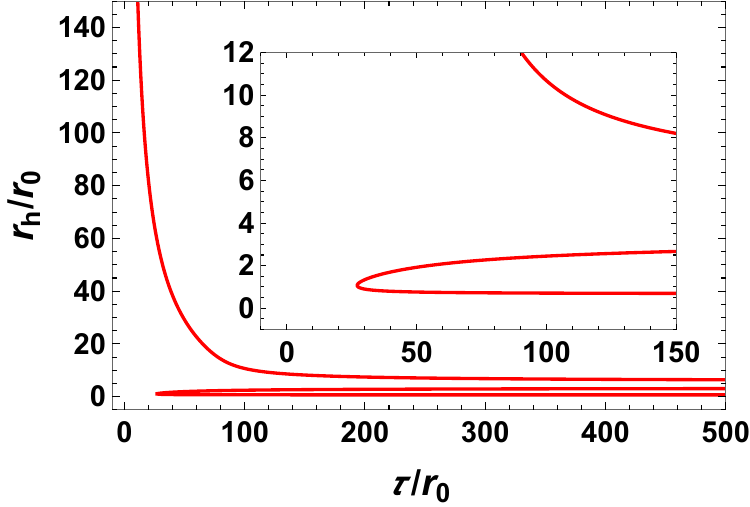}
\caption{Defect curves in the $r_h/r_0-\tau/r_0$ plane with $P r_0^2$=0.0003. Obviously, there are two defect curves.}\label{pCBtaur}
\end{figure}

One wonders whether such topological number is independent of $\tau$. For the purpose, we show the defect curve in Fig. \ref{pCBtaur}. For $\tau/r_0<27.0802$, there is only one black hole state corresponding to a defect with positive number. However when $\tau/r_0>27.0802$, one new pair branches with opposite winding numbers emerge. Nevertheless, the topological number $W$ always remains unchanged. So we can arrive the conclusion that the black hole system for $P r_0^2$=0.0003 still holds $W=+1$, and it is in the same topological class as the case I.

On the other hand, quite different from the first case, we observe two defect curves. The top curve starts at $\tau/r_0=0$ with infinite $r_{h}/r_{0}$ and extends to $\tau/r_0=\infty$ keeping $r_{h}/r_{0}$ with a finite value. Both ends of the bottom defect curve are at $\tau/r_0=\infty$. Despite such novel characteristic pattern with two curves, the topological number keeps a constant. Meanwhile, we have at most three black hole states for a given $\tau/r_0$.

\subsection{Case III: $P r_{0}^{2}$=0.00004}

Here we would like to further decrease the pressure and to examine whether other new topological pattern can be observed. For simplicity, we take $P r_{0}^{2}$=0.00004 as an example.

\begin{figure}
	\centering{
	\subfigure[]{\includegraphics[width=5cm]{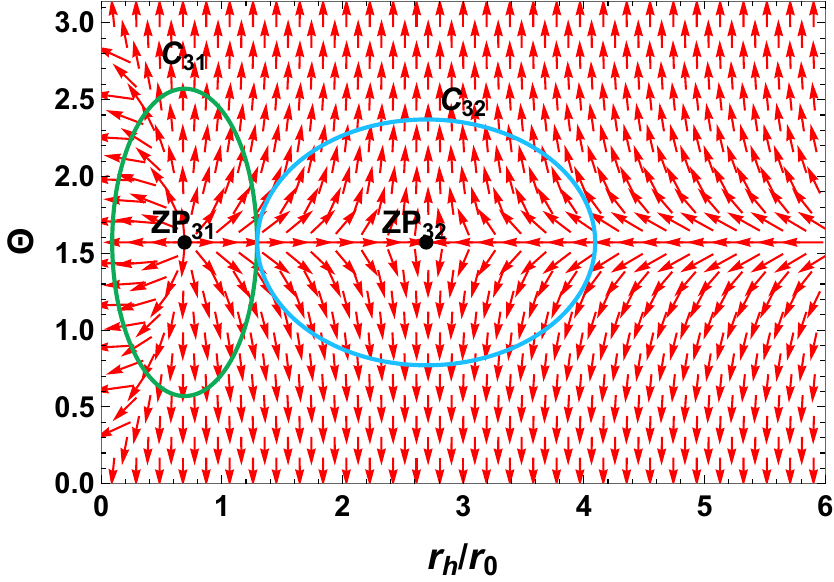}\label{CCvectora}}
	\subfigure[]{\includegraphics[width=5cm]{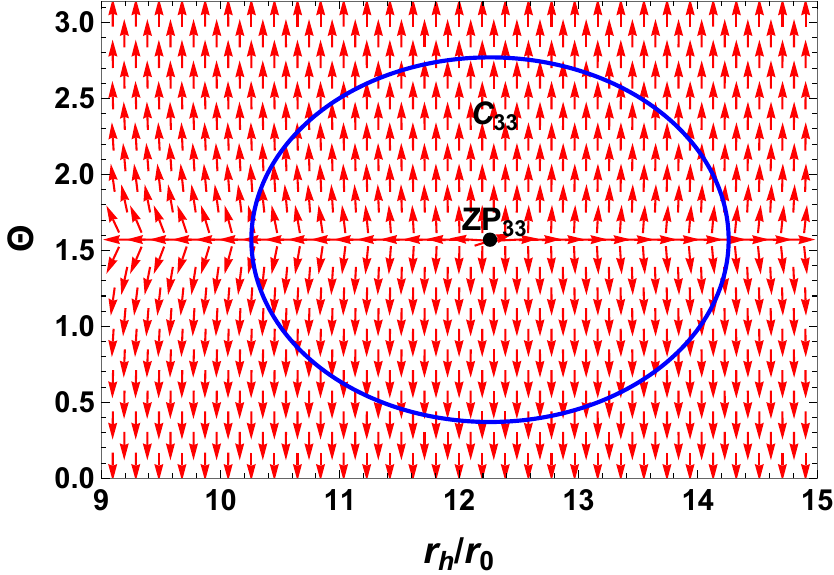}\label{CCvectorb}}\\
    \subfigure[]{\includegraphics[width=5cm]{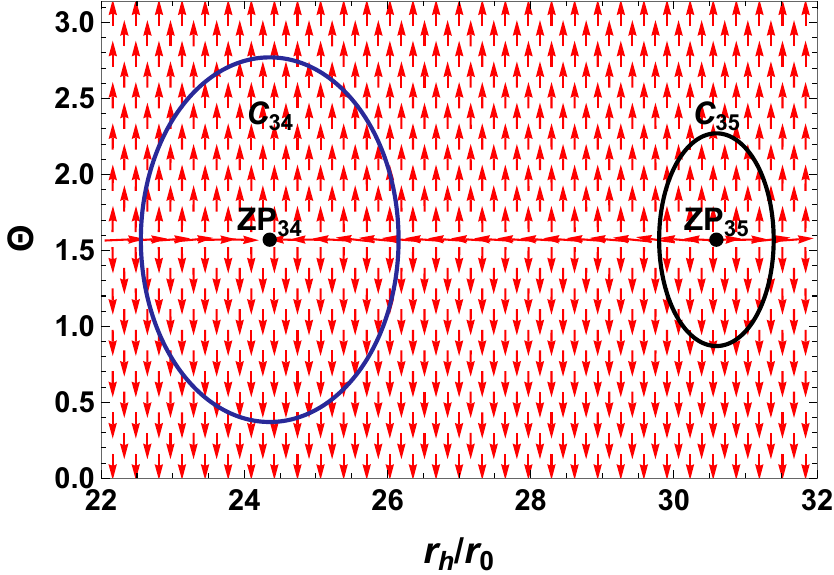}\label{CCvectorc}}
    \subfigure[]{\includegraphics[width=5cm]{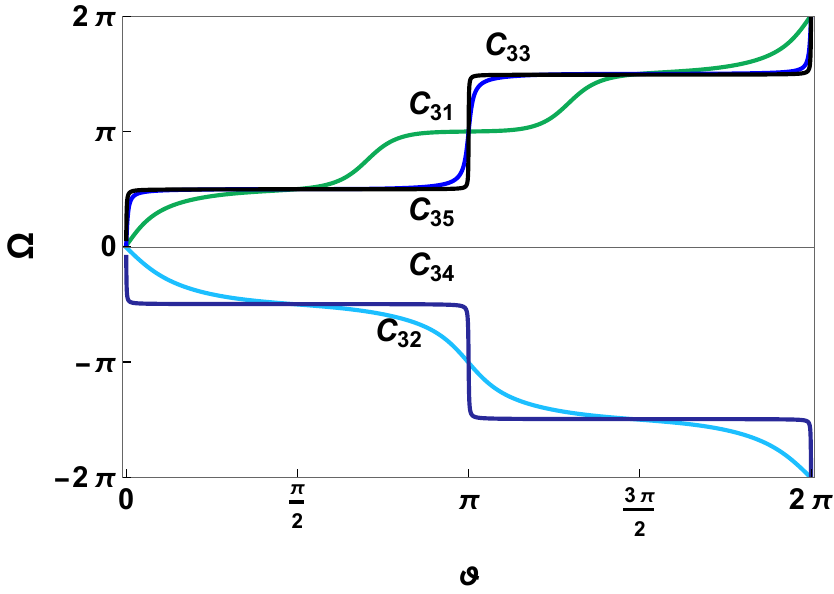}\label{CCtnumber}}}
	\caption{The vector unit field on the $r_h/r_0$-$\Theta$ plane with the $P r_0^2$ = 0.00004 and $\tau/r_0$ = 205. The zero points $ZP_{31}$, $ZP_{32}$, $ZP_{33}$, $ZP_{34}$, and $ZP_{35}$ located at ($r_{h}/r_{0}$, $\Theta$)=($0.6913$, $\pi/2$), ($2.6936$, $\pi/2$), ($12.2586$, $\pi/2$), ($24.3539$, $\pi/2$), and ($30.5935$, $\pi/2$) which are marked with black dots, and they are enclosed by the loops $C_{31}$, $C_{32}$, $C_{33}$, $C_{34}$, and $C_{35}$, respectively. (a) Small $r_{h}/r_{0}$ region. (b) Intermediate $r_{h}/r_{0}$ region. (c) Large $r_{h}/r_{0}$ region. (d) The deflection angle $\Omega$ as a function of $\vartheta$ for contours $C_{31}$ (green), $C_{32}$ (cyan), $C_{33}$ (blue), $C_{34}$ (dark blue), and $C_{35}$ (black).}\label{ppCCvectorc}
\end{figure}

Setting $\tau/r_0=205$, we illustrate the unit vector and the deflection angle in Fig. \ref{ppCCvectorc}. From the behavior of the vector shown in Figs. \ref{CCvectora}-\ref{CCvectorc}, we see there are five zero points $ZP_{3i}$ ($i$=1-5) located at $r_h/r_0$=0.6913, 2.6936, 12.2586, 24.3539, and 30.5935, respectively. The loops encircling these zero points are also parameterized with the form Eq. (\ref{pfs}) and with the parameters $(a, b, r_c)$=(0.6, 1, 0.6913), (1.4, 0.8, 2.6936), (2, 1.2, 12.2586), (1.8, 1.2, 24.3539), and (0.8, 0.7, 30.5935), respectively. By calculating the defection angle, we show the $\Omega(\vartheta)$ in Fig. \ref{CCtnumber}. From its behavior, we can immediately read the winding number of these zero points, i.e., $w_{1,3,5}=+1$ and $w_{2,4}=-1$. As a result, the topological number $W=\sum_{i=1}^{5}w_{i}=+1$, which remains the smae values obtained in the first two cases.

\begin{figure}
	\centering{
		\subfigure[]{\includegraphics[width=6cm]{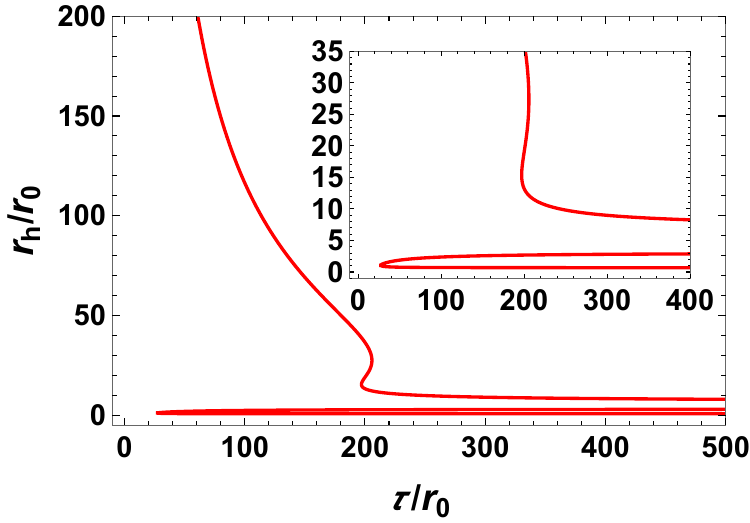}\label{CCtau}}}
	\caption{Defect curves in the $r_h/r_0-\tau/r_0$ plane with $P r_0^2$ = 0.00004.}\label{ppCCtau}
\end{figure}

Moreover, we present the defect curve in $r_{h}/r_{0}-\tau/r_{0}$ plane in Fig. \ref{ppCCtau}. Similar to the second case, it has two separated curves. The bottom curve share the similar behavior, however the top curve exhibits a non-monotonic behavior. Nevertheless, it does not affect the topological number.

\section{Discussion and conclusion}\label{Sec_Disscusion}

In this paper, we considered the topology by treating the dyonic black holes as defects in the thermodynamical parameter space. Different from previous black hole cases, the defect curves here can be one single, two separated curves. Despite the presence of such novel pattern of the multiple defect curves, our result confirms that the total topological number remains the same as that of the charged AdS black holes.

Since the dyonic black holes exhibit a new phase structure, we firstly considered it when two separated coexistence curves appear. Previous result given in Ref. \cite{qtLMDa} showed that there exists one no phase transition region for middle pressure. In this paper, we reconsidered it. By adjusting the coupling parameter $\alpha_1$, we found that such region could disappear. In particular, when $\alpha_1=0.41$, there exist both the intermediate-large and small-large black hole phase transition for the same pressure while with different temperature. This uncovers a novel phase structure for the dyonic black holes. More importantly, such phase structure with two separated coexistence curves indicates that the defect curve of the thermodynamical topology will give the multiple defect curves phenomenon.

In order to study the influence of the multiple defect curves on the topology of black hole thermodynamics, we followed the approach in Ref. \cite{topWei} and constructed a well behaved topology by making use of the off-shell free energy. In this topology, each dyonic black hole corresponds one defect in the thermodynamical parameter space. Meanwhile, the defect locates exactly at the zero point of the constructed vector. Winding number with values +1 or -1 can be endowed with each black hole, and which can reflect the thermodynamically locally stable or unstable. Especially, summing these winding number shall give a topological number, with which different black hole systems can be divided into different topological classes.

After establishing the topology, we examined the topological property for the dyonic black holes by taking $\alpha_1=0.41$ and $\alpha_2=50$. Three cases, $Pr_{0}^{2}$=0.05, 0.0003, and 0.00004, are investigated. For the first case, we observe only one defect at given $\tau/r_0=2$. Further varying it, the number of the defect keeps unchanged. This indicates that there is only one defect curve. Of particular interest, these defects on this curve all have winding number +1, leading to the topological number $W=+1$. Therefore, this dyonic black hole is in the same topological class of the charged RN-AdS black holes \cite{topWei}.

For other two cases, we observed that more than one defect exist at given $\tau/r_0$. For examples, we have three and five defects at most for the second and third cases, respectively. More significantly, the defect set will not be only one single curve, while two separated curves instead. At first glance, such phenomenon may result a different topological number. However, after calculating the topological number, we found that all the cases give $W=+1$, and this result is independent of the pressure and temperature.

In summary, we observed a novel phenomenon, multiple defect curves, for the thermodynamical topology of the dyonic black holes. However, it does not affect the topological number, and thus all these dyonic black hole systems with or without the multiple defect curves are in the same topological class. Our results provide an important ingredient to the black hole thermodynamical topology, and more interesting phenomena are expected to be uncovered.

\section*{Acknowledgements}
This work was supported by the National Natural Science Foundation of China (Grants No. 12075103, No. 12247101).

\end{document}